# Optical Characteristics of Charge Carrier Transfer across Interfaces between $YBa_2Cu_3O_{6+\delta}$ and $La_{0.7}Ca_{0.3}MnO_3$

A. Seo,[1,2] A. V. Boris,[1] G. Cristiani,[1] H.-U. Habermeier,[1] and B. Keimer[1]

[1]*Max-Planck-Institut für Festkörperforschung, Heisenbergstraße 1, D-70569 Stuttgart, Germany*
[2] *Department of Physics and Astronomy, University of Kentucky, Lexington 40506, USA*

We report a spectral ellipsometry study of multilayers composed of superconducting $YBa_2Cu_3O_{6+\delta}$ (YBCO) and ferromagnetic $La_{0.7}Ca_{0.3}MnO_3$ (LCMO) in the spectral range 0.7 – 6.5 eV. With increasing YBCO sublayer thickness, the optical spectral weight is enhanced at photon energies of 1.5 – 3.5 eV. The spectral weight enhancement is proportional to the number of interfaces of each multilayer sample, indicating its association with the interfacial electronic structure. Based on calculations in the framework of a multilayer model, we find that the shape of the interface-induced spectral weight is consistent with transfer of hole-carriers from YBCO to LCMO. Our results imply that the holes that are transferred across the interfaces accumulate in the LCMO layers, rather than being pinned by interfacial defects or annihilated by electron donors such as oxygen vacancies. Optical spectroscopy can thus serve as a non-destructive probe of charge transfer across buried interfaces in metal-oxide heterostructures.

## I. INTRODUCTION

Heterostructures of metal oxides with strongly correlated electrons offer a powerful platform to realize novel electronic and magnetic states in condensed matter [1]. Interfaces of superconducting $YBa_2Cu_3O_{6+\delta}$ (YBCO) and ferromagnetic $La_{0.7}Ca_{0.3}MnO_3$ (LCMO) have been of particular interest. Originally conceived to study the interplay between superconductivity and ferromagnetism [2-4], YBCO-LCMO heterostructures have also served as model systems for magnetic and orbital reconstructions [5,6], phonon hybridization [7], and superconducting spintronics [8,9]. Recently, charge density waves (CDWs) have been observed [10] in YBCO-LCMO multilayers based on optimally doped YBCO ($\delta \sim 1$), which does not exhibit CDW order in bulk form [11]. These results indicate that YBCO-LCMO heterostructures are suitable as a platform for research on the interplay between superconductivity, CDW order, and other correlated-electron phases. However, the investigation of such interfacial phenomena is difficult because the heterointerfaces are buried inside the materials, and spectroscopic probes mostly provide volume-averaged information. For example, far-infrared spectroscopic ellipsometry shows a significant decrease in the Drude spectral weight in YBCO/LCMO multilayers compared to their bulk counterparts [12]. While this observation is consistent with the depletion of 0.2-0.3 holes per Cu ion from the YBCO sublayers [3,6,13], it has been difficult to probe experimentally whether the hole-carriers from the YBCO layers are transferred to the neighboring LCMO layers, pinned at the interface by defects, or annihilated by electron donors such as oxygen vacancies.

In this paper we show that optical spectroscopy with photon energies > 0.7 eV is complementary to far-infrared spectroscopy and provides experimental evidence of hole-carrier transfer from YBCO to LCMO sublayers with a length scale of multiple unit cells. Specifically, we have observed increased spectral weight in the optical conductivity spectra of the multilayers, compared to those of the bulk compounds, in the photon energy region 1.5 – 3.5 eV. By systematically varying the sublayer thickness of the multilayer samples, we found that the increased spectral weight is linearly proportional to the number of YBCO/LCMO heterointerfaces. Simulations using a multilayer model provide useful information on the length scale of the charge transfer and on the interfacial electrodynamic properties of the YBCO/LCMO multilayers.

## II. EXPERIMENTAL DETAILS

We have grown YBCO/LCMO multilayers by pulsed laser deposition on $SrTiO_3$ (001) single crystal substrates. Details of the synthesis conditions and characterization of the lattice structure and interfacial properties of YBCO/LCMO multilayers can be found in Refs. [7,12]. While maintaining the total thickness of the multilayer samples at ~200 nm, we varied the YBCO sublayer thickness from 5 to 40 nm with constant (10 nm) LCMO sublayer thickness, *i.e.* $(YBCO_x/LCMO_{10\,nm})\times y$ with $(x, y)$ = (5 nm, 14), (10 nm, 10), (15 nm, 8), (20 nm, 7), (25 nm, 6), and (40 nm, 4). The *dc* magnetic properties of the multilayers were measured using a magnetic property measurement system (MPMS, Quantum Design) consisting of a vibrating sample magnetometer with a superconducting quantum interference device. The magnetic field was applied parallel to the in-plane direction. The *ac*-magnetic susceptibilities of the samples were measured by using a physical property measurement system (PPMS, Quantum Design) by applying the magnetic field along the surface normal direction of the multilayers. Optical spectroscopic ellipsometry measurements were performed at room temperature using a Variable-Angle Spectroscopic Ellipsometer (VASE, Woollam) in the photon energy range 0.7 – 6.5 eV at incident angles of 60º, 70º, and 80º. Accurate values of the real and imaginary parts of the dielectric functions can be obtained from the ellipsometric angles $\Psi(\omega)$ and $\Delta(\omega)$ extracted from the optical spectra, without the need of a Kramers-Kronig analysis [14]. Note that the dielectric functions obtained in this way are consistent with all the spectra measured at different incident angles. Moreover, the probing depth of photons in this energy range is larger than 50 nm, which makes this technique useful for non-destructive measurements of the electronic properties of buried interfaces. The dielectric functions of the multilayers were analyzed using the WVASE software, which derives the solutions of the multiple Jones matrices using an isotropic-layer/semi-infinite-substrate model by a numerical iteration process [14]. The obtained dielectric functions reflect the in-plane optical spectra of the multilayers with better than 90 % accuracy, even though the uniaxial anisotropy (with reasonably different out-of-plane dielectric functions) is neglected, according to our full anisotropic spectral analyses.

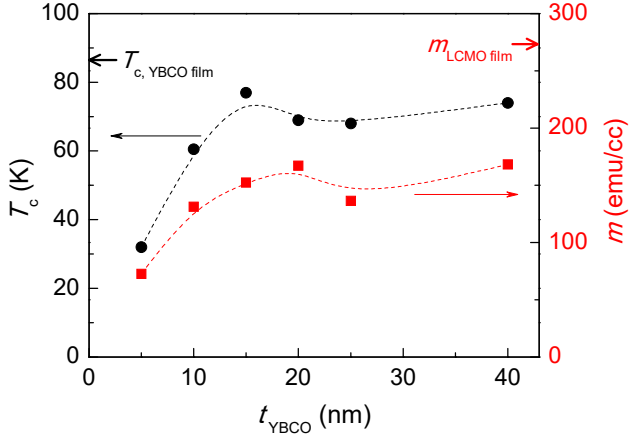

**FIG. 1.** Superconducting $T_c$ of YBCO/LCMO multilayers measured by *ac*-susceptibility, and magnetic moment $m$ at temperature 100 K measured in a magnetic field of 100 Oe, normalized by the volume of the LCMO sublayers. The dashed lines are guides to the eye.

## III. EXPERIMENTAL RESULTS

Figure 1 displays *ac*-susceptibility measurements of our YBCO/LCMO multilayer samples. The thickness dependences of the superconducting transition temperature ($T_c$) and the ferromagnetic moment ($m$) are closely parallel. Note that $T_c$ decreases as the thickness of YBCO decreases (YBCO thin films prepared under identical conditions show $T_c$ = 83 K). Figure 1 also shows $m$ measured at 100 K, normalized by the total volume of the LCMO sublayers. As the YBCO sublayer thickness decreases, $m$ also decreases. Thus, both superconductivity and ferromagnetism weaken simultaneously as the YBCO sublayer thickness decreases. This observation can be understood as a consequence of an interfacial electronic reconstruction, i.e. hole-carrier transfer from YBCO to LCMO sublayers, which reduces the density of superconducting carriers (holes) in a YBCO sublayer while promoting the LCMO sublayer into the charge-ordered, antiferromagnetic regime of the phase diagram [5,15]. The optical resonance and spectral features discussed below provide experimental evidence to support the hole-carrier transfer from YBCO to LCMO sublayers.

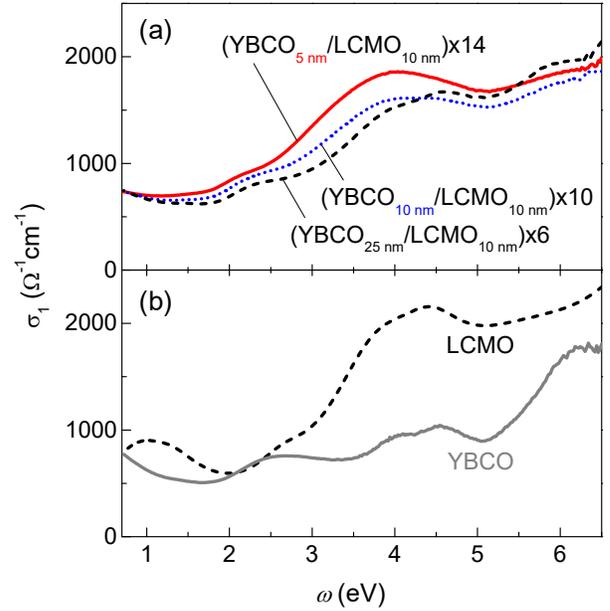

**FIG. 2. (a)** Room-temperature optical conductivity ($\sigma_1(\omega)$) spectra of YBCO/LCMO multilayers of thickness 200 nm. The YBCO sublayer thickness varies while the LCMO sublayer thickness is fixed to 10 nm. **(b)** $\sigma_1(\omega)$ reference spectra of YBCO and LCMO single-phase thin films (thickness about 100 nm).

Figure 2 displays the optical conductivity ($\sigma_1(\omega)$) of the multilayers, which increases in the 2 – 4 eV photon energy range as the YBCO sublayer thickness decreases. The figure also shows reference spectra of the YBCO and LCMO thin films, which are consistent with previously reported spectra [16-18]. The enhanced spectral weight in the 2 – 4 eV range is complementary to the reduced spectral weight observed by infrared spectroscopy at lower photon energies [12]. To analyze the spectral enhancement, we calculated the difference ($\Delta\sigma_1$) between an experimentally measured spectrum ($\sigma_{Exp}$) and a multilayer model spectrum ($\sigma_{Mod}$), i.e. $\Delta\sigma_1 = \sigma_{Exp} - \sigma_{Mod}$. Figure 3 (a) shows $\Delta\sigma_1$ and $\Delta\varepsilon_1$ of the (YBCO$_{10\text{ nm}}$/LCMO$_{10\text{ nm}}$)×10 multilayer sample. Note that there is a strong increase of $\Delta\varepsilon_1$ at lower energy due to the decrease in the concentration of conduction carriers in the multilayer, which is consistent with the infrared ellipsometry results of Ref. [12].

We have calculated the integrated optical spectral weight, $\Delta\text{sw} = \int \Delta\sigma_1 \cdot d\omega$, of the multilayers to check if the observed $\Delta\sigma_1$ is related to the heterointerfaces. Figure 4 shows $\Delta\text{sw}$ at 1.5 – 3.5 eV for all of our multilayer samples. Note that $\Delta\text{sw}$ gradually decreases as the YBCO sublayer thickness increases. By taking into account the number of interfaces of each multilayer within the probing depth (~70 nm) of the 1.5 – 3.5 eV photons, the normalized $\Delta\text{sw}$ (i.e. $\Delta\text{sw}$ per interface) shows a constant value of approximately 100 $\Omega^{-1}$cm$^{-1}$eV for all of our samples. This result indicates that the non-zero $\Delta\sigma_1$ is a characteristic feature of the optical properties of the heterointerfaces between YBCO and LCMO.

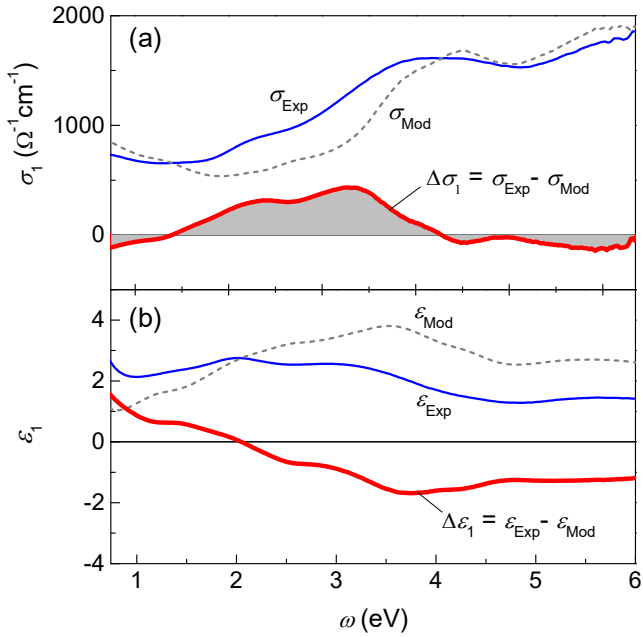

**FIG. 3.** (a) $\sigma_1(\omega)$ and (b) $\varepsilon_1(\omega)$ spectra of a YBCO$_{10\,nm}$/LCMO$_{10\,nm}$ multilayer (blue solid lines), and the multilayer model spectra calculated with the stacking sequences realized in our samples (dashed lines). The red solid line shows the difference between the experimental spectra and the multilayer model spectra.

Based on our experimental results above, we have simulated $\Delta\sigma_1$ to understand the electronic reconstruction at the heterointerfaces between YBCO and LCMO. This spectral simulation carries some uncertainty due to the complexity of the heterointerfaces, where multiple ions are involved, and the lack of good reference spectra of these interfaces. We therefore compare the experimentally obtained $\Delta\sigma_1$ with several possible spectral simulations using modified dielectric functions. To simplify the process, we assume that the dielectric functions of the interior of both YBCO and LCMO sublayers remain the same as those of the reference thin films.

Following prior work on YBCO-LCMO heterostructures [5,15], we assume that the near-interface region consists of an electron-depleted layer where the Mn ions are predominantly in the valence state 4+, as shown in the schematic diagram of Fig. 5 (b). When the Mn valence increases from 3+ to 4+, we expect that the Fermi-level shifts downward in the Mn $3d$ band. This assumption is consistent with the peak shift of the O $2p$ to Mn $3d$ charge-transfer transitions from ~4.4 eV (LCMO) to ~3.6 eV (CaMnO$_3$) [19]. Hence, for the LCMO interface region, we have used a modified LCMO dielectric function where the charge transfer transition is lowered by 1 eV from the LCMO spectra, while fixing the other optical transitions. The simulated $\Delta\sigma_1$ (Fig. 5 (b)) accounts for much of the experimentally measured spectral-weight modification (Fig. 5 (a)), and it gradually increases as we increase the thickness of the interfacial LCMO layer with Mn$^{4+}$ ions. We have simulated a few other scenarios such as hole-carrier transfer from LCMO to YBCO sublayers, which creates a Mn$^{3+}$-rich layer at the interface, but none of these simulated spectra agreed as closely with the experiment as the one above.

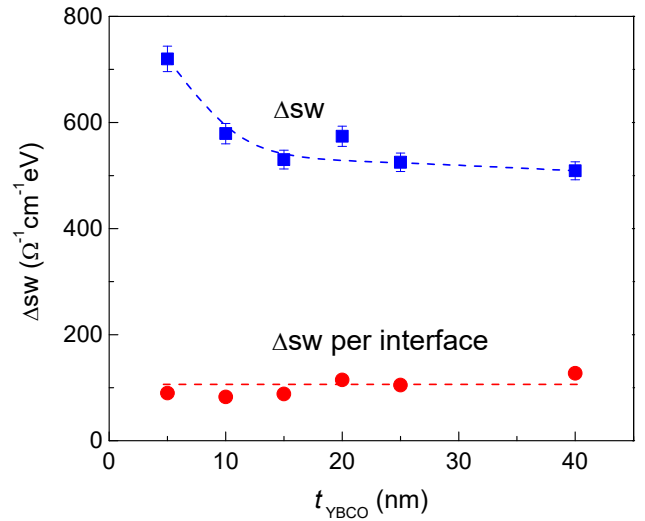

**FIG. 4.** Integrated spectral weight of $\Delta\sigma_1$ at 1.5 – 3.5 eV as a function of YBCO sublayer thickness (squares). Circles indicate the integrated spectral weight of $\Delta\sigma_1$ normalized by the number of interfaces within the probing depth of the photons (~70 nm). The dashed lines are guides to the eye.

## IV. DISCUSSION

In the simulations displayed in Fig. 5, we only modified the optical response of the LCMO layers near the interface, leaving the one of the YBCO layered unchanged. The good overall agreement with the experimental data implies that changes in the electronic structure of the LCMO sublayers dominates the $\Delta\sigma_1$ spectra in the energy range 1.5 – 3.5 eV. Conversely, the infrared spectra (< 0.7 eV) are strongly affected by the reduction of the Drude response in the interfacial regions of the YBCO sublayers due to the transfer of mobile hole to the LCMO layers [12].

It is worthy to note that $\Delta\sigma_1$ in Fig. 5 (a) has two peaks at 2.4 and 3.6 eV. While the intense peak at 3.6 eV can be understood as a fingerprint of an electron-depleted region at the LCMO interface, the weaker peak at 2.4 eV is not reproduced by the simulation. In this energy range, a prior optical study of YBCO thin films revealed a 1.4 eV feature in the optical conductivity, which was assigned to a charge-transfer transition from the O $2p$ band to the Cu $3d$ band in the CuO$_2$ planes [20], as well as a 2.6 eV feature, which was attributed to an excitation from a manifold of initial states into the anti-bonding Cu(2)-O(2)-O(3) band [21]. Hence, the peak at 2.4 eV may be related to changes of the electronic states in the YBCO sublayers that are difficult to reproduce in optical simulations based on existing bulk dielectric functions. More elaborate calculations are required to properly account for the spectral-weight rearrangement in the YBCO layers. Leaving these details aside, we note that none of our optical spectra exhibit a sharp peak at 4 eV which is related with oxygen vacancies in YBCO, as observed in highly oxygen-deficient YBa$_2$Cu$_3$O$_{6.1}$ bulk crystals [22]. This finding suggests that oxygen vacancies do not play an important role in this multilayer system, in agreement with recent measurements of the CDW wavevector in YBCO-LCMO multilayers [10].

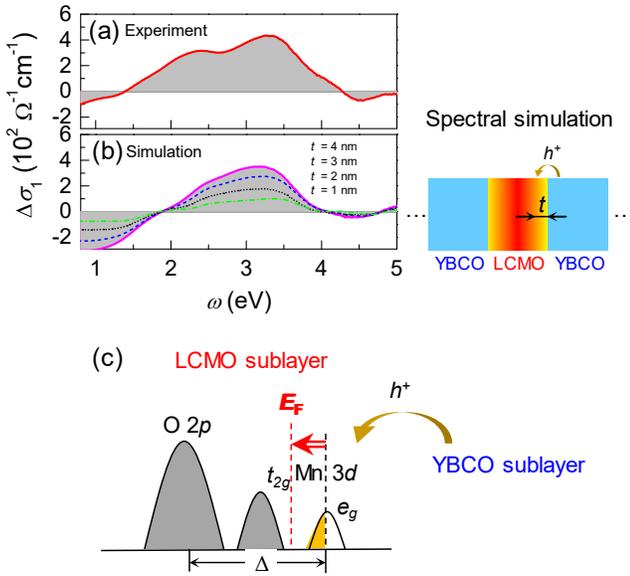

**FIG. 5.** (a) Experimental $\Delta\sigma_1$ of a YBCO$_{10\,nm}$/LCMO$_{10\,nm}$ multilayer. (b) Simulated $\Delta\sigma_1$, as described in the text. Mn ions in the 4+ valence state are assumed to be present in a layer of thickness $t$ around the LCMO/YBCO interface layer, (c) Schematic band diagram of the interfacial regions of the LCMO sublayers showing the hole-carrier transfer from the YBCO sublayers and the reduced Fermi level.

It is reasonable to conclude that the YBCO sublayer is depleted of holes (i.e. superconducting carriers at low temperatures) and the holes are transferred to the adjacent LCMO sublayers, as schematically shown in Fig. 5 (c). Hence, antiferromagnetic superexchange interactions between Mn ions can prevail over ferromagnetic double-exchange interactions in LCMO. As a result, the ferromagnetic moment per LCMO volume is reduced as the hole carriers are transferred from YBCO sublayers. At the same time, the density of superconducting carriers in YBCO is reduced, which explains the parallel behavior between the ferromagnetic moment of LCMO and the superconducting $T_c$ of YBCO in these multilayers, as shown in Fig. 1.

This conclusion is in line with prior x-ray [5] and scanning tunneling spectroscopy [15] experiments as well as ab-initio calculations [13] on YBCO-LCMO interfaces. Further support for hole transfer from YBCO to LCMO comes from Raman spectroscopy and resonant x-ray scattering experiments that have revealed that the electron-phonon coupling [7] and the formation of CDWs [1] in a YBCO sublayer can be understood consistently within this scenario. On the other hand, electron energy loss spectroscopy results indicating that the electron density in the LCMO layer increases as the interface is approached disagree with our results [23].

## V. CONCLUSION

In summary, we have observed a spectral resonance effect in YBCO/LCMO multilayers using optical spectroscopic ellipsometry at photon energies in the range 0.7 – 6.5 eV. Since the optical resonance strength is proportional to the number of the interfaces in each multilayer, its spectral shape can be regarded as a fingerprint of the electronic state at and near the interfaces. Multilayer spectral simulations provide an explanation of its salient features in terms of a hole-carrier transfer from YBCO to LCMO sublayers across the hetero-interfaces. This electronic reconstruction at the interface provides a key to understanding the parallel suppression of superconductivity and ferromagnetism of YBCO/LCMO multilayers. Optical spectroscopy, in conjunction with simulations, can thus serve as a highly specific probe of interfacial reconstructions in metal-oxide heterostructures and multilayers.

**ACKNOWLEDGEMENT**

We thank G. Khaliullin, V. Hinkov, and S. Okamoto for valuable discussions and Y. Matiks, P. Popovich, B. Bruha, H. Uhlig, and M. Schulz for assisting with experimental work. A.S. acknowledges support (Research Fellowship for Experienced Researchers) from the Alexander von Humboldt Foundation and the National Science Foundation Grant No. DMR-1454200. B.K. acknowledges support of the DFG under grant TRR80.